# Quality assurance of the Federal Interagency Traumatic Brain Injury Research (FITBIR) database for multi-site MRI analysis

Adam M. Saunders*[a], Michael E. Kim[b], Gaurav Rudravaram[a], Elyssa M. McMaster[a], Chloe Scholten[c,d], Sequoia Wade[e], Marselle Rasdall[e], Simon Vandekar[f], Tonia S. Rex[e,g], François Rheault[h], Bennett A. Landman[a,b,g,i]

[a]Department of Electrical and Computer Engineering, Vanderbilt University, Nashville, TN, USA; [b]Department of Computer Science, Vanderbilt University, Nashville, TN, USA; [c]Hotchkiss Brain Institute, University of Calgary, Calgary, AB, Canada; [d]Alberta Children's Hospital Research Institute, University of Calgary, Calgary, AB, Canada; [e]Department of Ophthalmology and Visual Sciences, Vanderbilt University Medical Center, Nashville, TN, USA; [f]Department of Biostatistics, Vanderbilt University Medical Center, Nashville, TN, USA; [g]Department of Radiology and Radiological Science, Vanderbilt University Medical Center, Nashville, TN, USA; [h]Computer Science Department, Université de Sherbrooke, Sherbrooke, QC, Canada; [i]Department of Biomedical Engineering, Vanderbilt University, Nashville, TN, USA

## ABSTRACT

The Federal Interagency Traumatic Brain Injury Research (FITBIR) database is a centralized data repository for traumatic brain injury (TBI) research. It includes over 45,529 magnetic resonance images (MRI) from 6,211 subjects (9,229 imaging sessions) across 26 studies with heterogeneous organization formats, contrasts, acquisition parameters, and demographics. In this work, we organized and harmonized all available structural and diffusion MRI from FITBIR along with relevant demographic information into the Brain Imaging Data Structure. We analyzed whole-brain mean fractional anisotropy, mean diffusivity, total intracranial volume, and the volumes of 132 regions of interest using UNesT segmentations. There were 4,868 subjects (7,035 sessions) with structural MRI and 2,666 subjects (3,763 sessions) with diffusion MRI following quality assurance and harmonization. We modeled profiles for these metrics across ages with generalized additive models for location, scale, and shape (GAMLSS) and found significant differences in subjects with TBI compared to controls in volumes of 15 regions of the brain ($q < 0.05$, likelihood ratio test with false discovery rate correction).

## 1. INTRODUCTION

Traumatic brain injury (TBI) encompasses a wide variety of presentations and outcomes, often classified into mild, moderate and severe based on the Glasgow coma scale[1]. The Glasgow coma scale is unevenly correlated with outcomes and does not capture the full patient experience[2]. Coarse ratings can obscure subgroups of TBI presentation, leading to efforts to characterize TBI on sensitive and multi-dimensional frameworks such as clinical presentation, blood biomarkers, imaging, and modifying factors (CBI-M)[3].

Structural and diffusion-weighted magnetic resonance imaging (MRI) provide insight into volumetric and microstructural changes in the brain following TBI. Even in mild TBI, the brain undergoes volumetric changes that correlate with outcomes[4]. Diffusion tensor metrics like fractional anisotropy (FA) and mean diffusivity (MD) are predictive of TBI outcomes[5].

*adam.m.saunders@vanderbilt.edu

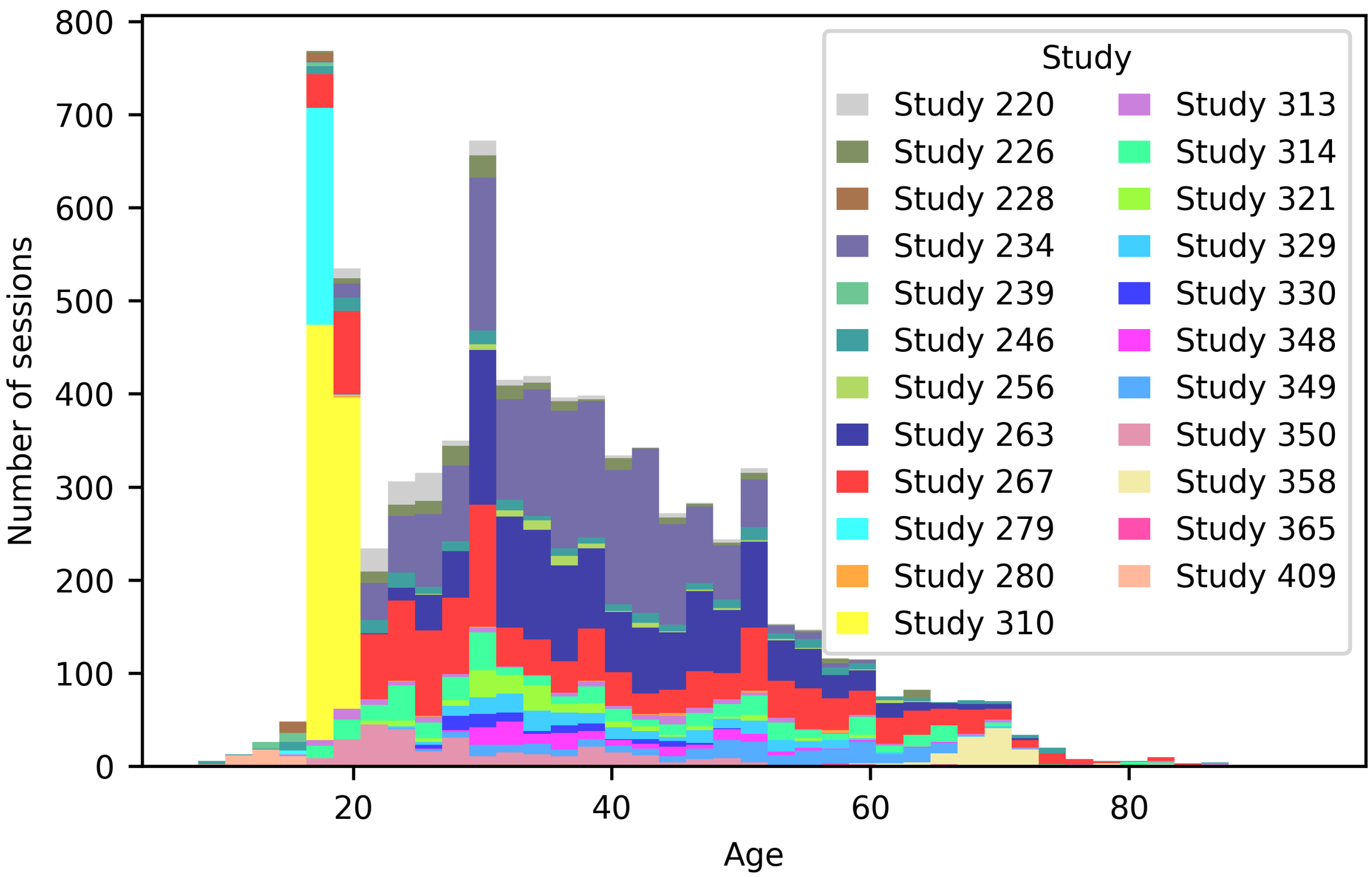

Figure 1. We unify 6,211 subjects (9,229 sessions) across 26 studies from the Federal Interagency Traumatic Brain Injury Research (FITBIR) database by organizing the images and relevant demographics and metadata into BIDS format. (Only sessions from studies with age reported are shown here.)

The search for more sensitive classification systems for TBI severity has highlighted the need for large-scale data repositories like the Federal Interagency Traumatic Brain Injury Research (FITBIR) database in the United States[6] and the Collaborative European NeuroTrauma Effectiveness Research in TBI (CENTER-TBI) database[7]. We aim to analyze diffusion tensor metrics and brain region volumes across all MRI available from FITBIR (Fig. 1). In this work, we show how we can use two harmonization methods (HACA3 and ComBAT) to curate a useful subset of structural and diffusion MRI from FITBIR for analysis. We demonstrate significant differences in the volumes of 15 brain regions among patients with TBI using generalized additive models for location, scale, and shape (GAMLSS).

## 2. METHODS

We retrieved all available structural and diffusion MRI from FITBIR. We converted all images to NIfTI format with dcm2niix[8] and reorganized available data into the Brain Imaging Data Structure (BIDS)[9]. We recorded common data elements in each study to store age, sex, race, ethnicity, case/control status, Glasgow coma scale score, scanner type, scanner ID, and location of scanner as available.

Next, we visualized a montage of slices from each image and performed quality assurance with AutoQA[10]. We removed images with 2D slices, extremely limited fields of view, and severe anatomical deformations or artifacts. We included studies containing participants with a reported age, sex, and case/control status and sessions with diffusion MRI or a T1-weighted image available (Fig. 2).

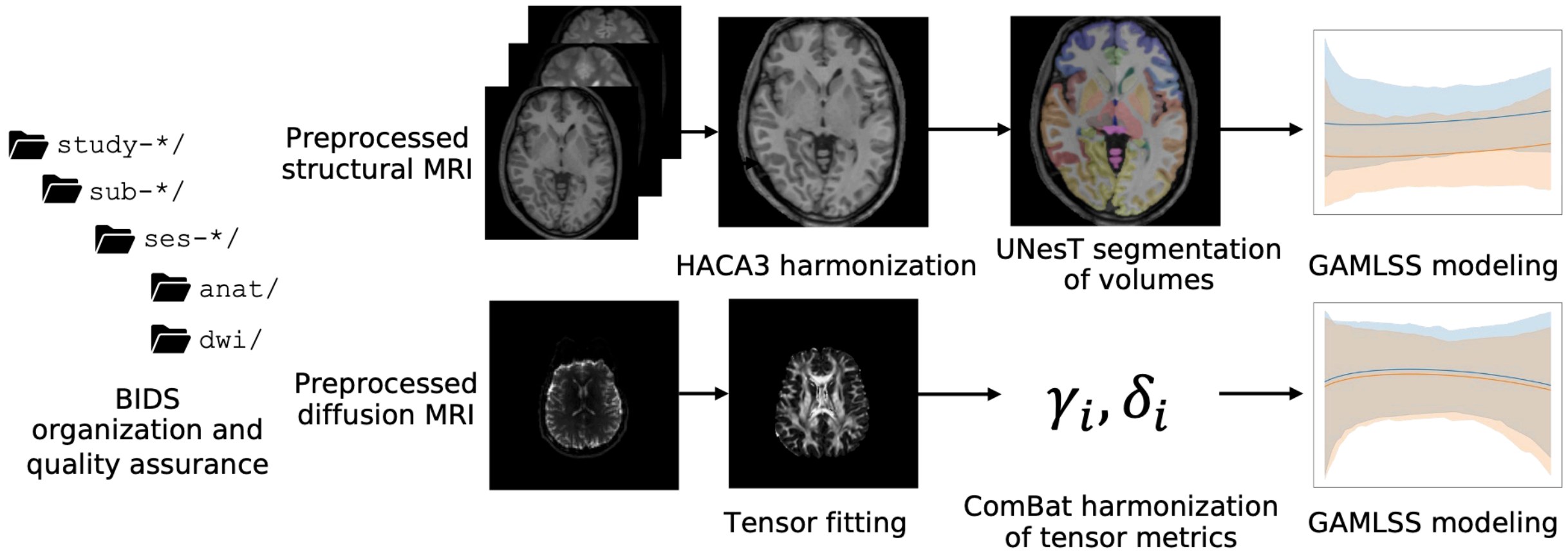

Figure 2. We harmonized the structural MRI using HACA3 and calculated the volumes of 132 regions of interest. For the diffusion MRI, we first calculated whole-brain mean FA, mean MD, and total intracranial volume (TICV) and harmonized these metrics with ComBat. We used generalized additive models for location, scale, and shape (GAMLSS) to model these metrics across ages 15-90 years in patients with TBI versus controls.

For the structural MRI, we harmonized the images as described in Saunders et al.[11]. We used the harmonization with attention-based contrast, anatomy, and artifact awareness (HACA3) framework. HACA3 is a deep learning harmonization framework that fuses information from available contrasts to produce an image with the same contrast as a target image[12]. By fusing available contrasts, we are able to incorporate all available imaging and produce images with a unified contrast. For preprocessing, we used SMORE super-resolution[13], N4 bias correction[14], and rigid registration to MNI space. We applied HACA3 harmonization to fuse information from all structural MRI and harmonize contrasts to a single T1-weighted image[12]. We calculated the volumes for 132 regions of interest (ROIs) using UNesT[15].

We preprocessed all diffusion MRI sessions using PreQual, which includes denoising, normalization, susceptibility and eddy current nonlinearity correction[16]. We combined multiple acquisitions with separately acquired $b_0$ images or reverse phase encoding images where possible. In acquisitions with different echo times per shell, we only used the shell nearest to a b-value of 1000 s/mm$^2$.

To process the diffusion MRI, we used nf-neuro, a set of neuroimaging pipelines and workflows based on the Nextflow workflow manager[17]. We skull-stripped the preprocessed T1-weighted images with SynthStrip[18] and applied rigid registration to the diffusion image. For the diffusion MRI, we resampled to 1 mm isotropic resolution, skull-stripped using the mean $b_0$ image, corrected for free water contributions to the diffusion signal[19], and fit tensors to the signal. As the diffusion tensor model assumes Gaussian diffusion, we only analyzed sessions that have at least 12 b-values between 500 s/mm$^2$ and 1500 s/mm$^2$, and we ignored volumes that only included higher b-values. We calculated TICV, mean FA, and mean MD within all brain tissue (excluding ventricles). Because HACA3 was developed for structural MRI harmonization and cannot readily harmonize 4D diffusion MRI, we harmonized these resulting metrics using ComBat. ComBat is a method for statistical harmonization of metrics by modeling and estimating additive and multiplicative bias from the data[20]. Unlike HACA3, ComBat requires an explicit mapping of which data belong to each site. Here, we define the study as site along with age, sex, and case/control status as covariates[21]. We ran quality assurance after all processing with AutoQA.

## 3. RESULTS

A total of 4,868 subjects (7,035 sessions, 19 studies) passed inclusion criteria and quality assurance for structural MRI, while there were 2,666 subjects (3,763 sessions, 13 studies) with diffusion MRI and a corresponding T1-weighted image.

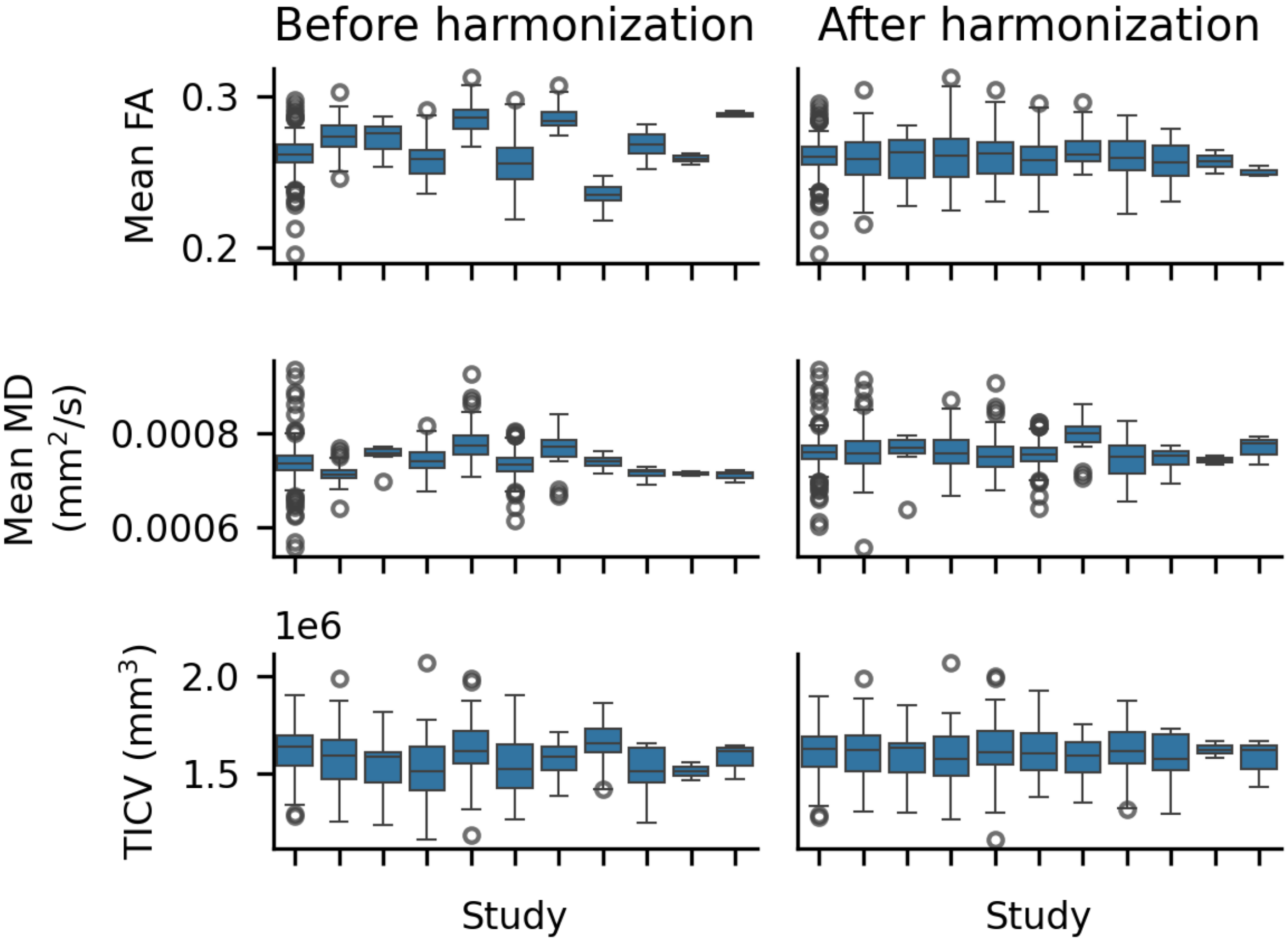

Figure 3. There are significant differences in metrics even among the controls from FITBIR studies with age, sex, and case/control status reported ($p < 0.05$ for each metric, ANOVA). ComBAT reduces biases across studies ($p < 0.05$ for mean MD only).

To remove potential covariate bias between groups, we used propensity score matching[22] based on age and sex to match controls and patients with TBI. After matching, there were 1,982 subjects with structural MRI and 1,060 subjects with diffusion MRI. Before harmonization, the mean FA, mean MD, and TICV demonstrated significant differences ($p < 0.05$ using ANOVA. ComBat reduced the biases among sites, with only 1 metric having significant differences after harmonization (Fig. 3).

To describe trends across age in these metrics in patients with TBI and controls, we fit GAMLSS curves to the ROI volumes and diffusion tensor metrics based on age, sex and case/control status using the generalized Gamma distribution family[23]. We rejected outliers more than five standard deviations from the mean value for each study (296 measurements from structural MRI and 18 measurements from diffusion MRI). Upon visual inspection, many outliers were due to errors in skull-stripping and segmentation.

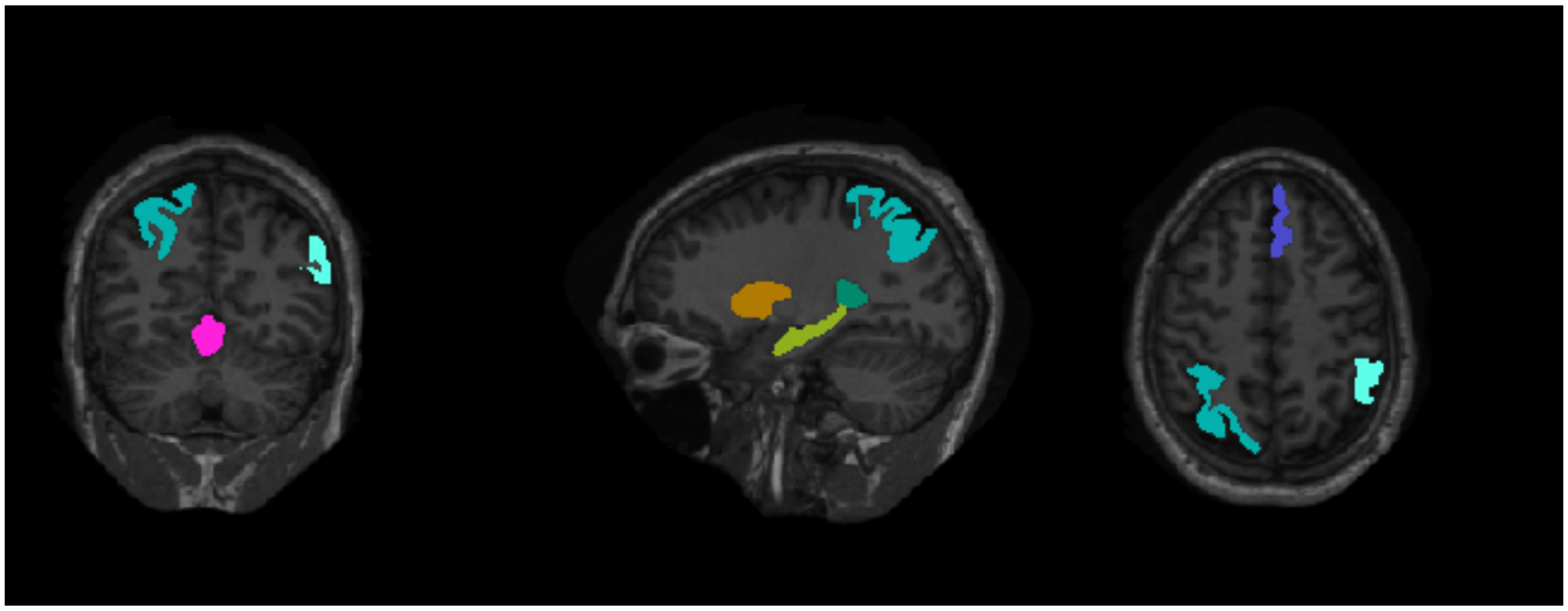

Figure 4. 15 out of 132 ROI volumes have significant differences, including both hippocampi and several cortical regions that were not symmetric across the brain ($q < 0.05$, likelihood ratio test with FDR correction).

We bootstrapped 95th percentile confidence intervals and ran a likelihood ratio test using a null model of curves fit without case/control status. We used Benjamini-Hochberg multiple comparisons correction with a false discovery rate (FDR) of 0.05. We found no significant difference in the models for whole-brain mean FA, mean MD, and TICV, whereas there were significant differences in 15 out of 132 ROIs from structural MRI, including both hippocampi and gyri recti, and parts of the insular cortex and ventricles. With the exception of the hippocampi and gyri recti, the significant regions are largely not symmetric across the hemispheres of the brain (Fig. 4). In the regions with significant differences, ventricle size increased and brain tissue decreased in volume in those with TBI (Fig. 5). Running the analysis using only studies that collected data from controls did not affect the results, indicating that there were not major biases among the studies without controls that could have substantially affected the results. In summary, 1) harmonization significantly reduced differences among the controls from each FITBIR study, and 2) we identified significant differences in the volumes of 15 regions in patients with TBI.

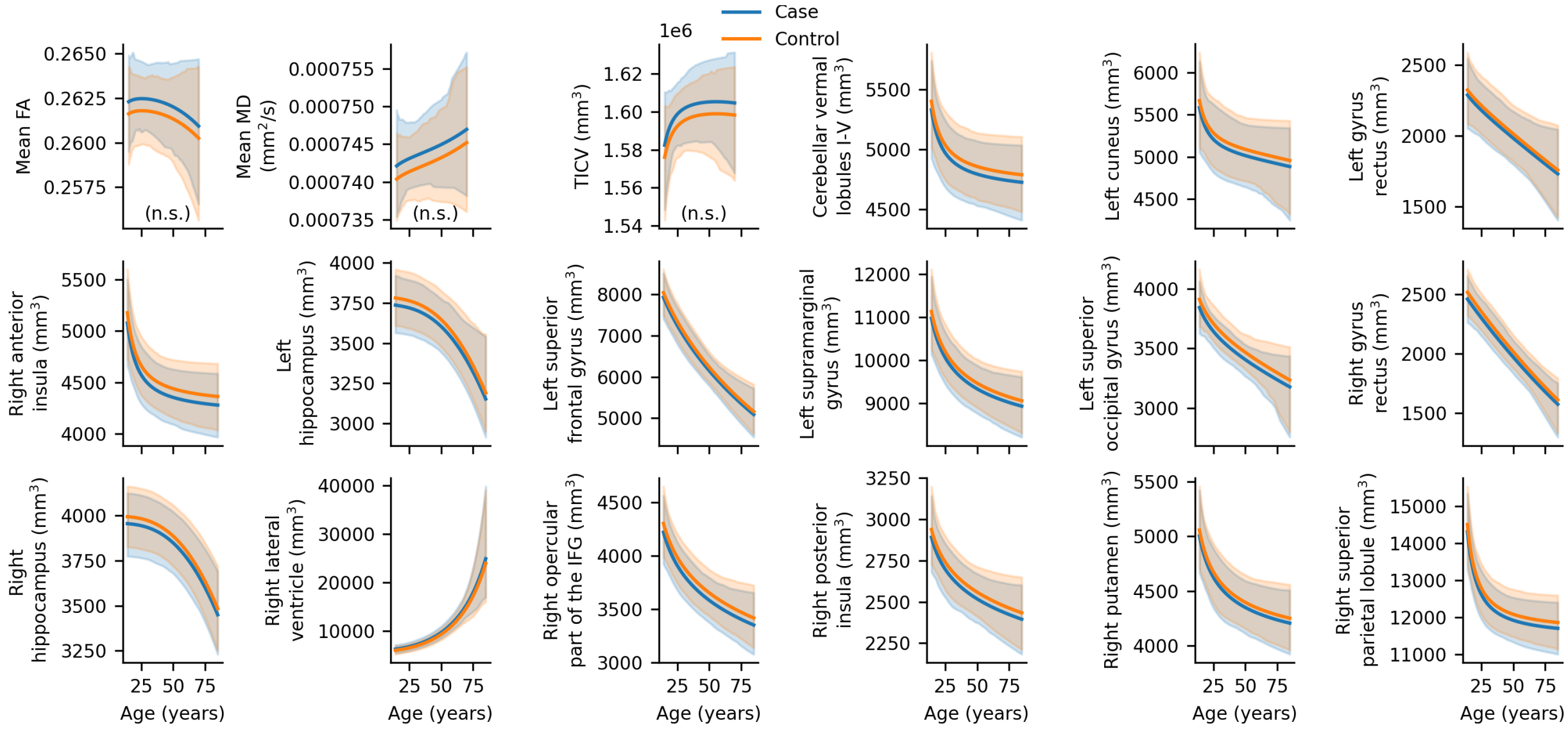

Figure 5. Mean FA, mean MD, and TICV were not significantly different between TBI and controls. The 15 out of 132 ROI volumes that have significant differences vary in terms of trajectory across ages. Overall, ventricle size and brain tissue decreased in volume in those with TBI ($q < 0.05$, likelihood ratio test with FDR correction).

## 4. DISCUSSION

The use of large-scale retrospective structural and diffusion MRI in this study highlights the need for consistent data collection, organization, and processing, especially with heterogeneous pathologies. Only 77% of structural MRI sessions and 62% of diffusion MRI sessions were suitable for analysis, largely due to missing demographic information such as age, sex or case/control status (~10% of all sessions) and poor image quality or severe anatomical deformations (~10% of structural and ~20% of diffusion sessions). Several studies had missing diffusion gradient information that was unrecoverable (315 sessions) or T1-weighted images collected days apart from diffusion MRI, rendering it difficult to link information from multiple contrasts (81 sessions). A subtype analysis based on TBI severity using Glasgow coma scale would be useful. However, only three studies reported age, sex, and Glasgow coma scale, and these studies skewed very heavily toward mild TBI ($n = 19$ subjects with moderate or severe TBI).

Harmonization helped account for the large intra-study variability in derived metrics. HACA3 allowed for harmonization of structural MRI without an explicit definition of site, but it has not been routinely applied or validated for raw diffusion MRI. Instead, we performed harmonization of diffusion MRI on derived metrics with ComBat and analyzed these metrics independent of the structural MRI. Along with other considerations for careful distribution matching, ComBat requires an explicit mapping of site[24,25]. While we considered each study as a site, FITBIR contains several consortium-level studies with data from many institutions, and many institutions within FITBIR did not collect data from healthy controls as a reference.

## 5. CONCLUSION

We can curate a useful subset of data from FITBIR for harmonization and further analysis. Careful organization, processing, and quality assurance can account for variability in data quality, while harmonization can account for inter-site variation. TBI is an individualized condition, with large variations in injury mechanisms, severity, and patient experience. Despite this heterogeneity, our findings demonstrate how we can detect population-level changes in brain volume from large-scale imaging. With large-scale imaging available for harmonization and analysis from repositories like FITBIR, we can investigate complex and varied conditions like TBI with high statistical power.

## ACKNOWLEDGEMENTS

This work was supported by DoD grant HT94252410563, NIH 1R01EB017230, NIH P50HD103537, the Alzheimer's Disease Sequencing Project Phenotype Harmonization Consortium that is funded by NIA (U24 AG074855, U01 AG068057 and R01 AG059716), the Canadian Institutes of Health Research, the Canadian Neurodevelopmental Research Training Platform, and the Hotchkiss Brain Institute. The Vanderbilt Institute for Clinical and Translational Research is funded by the National Center for Advancing Translational Sciences Clinical Translational Science Award Program, Award Number 5UL1TR002243-03. This work was conducted in part using the resources of the Advanced Computing Center for Research and Education at Vanderbilt University, Nashville, TN. Data and/or research tools used in the preparation of this manuscript were obtained and analyzed from the controlled access datasets distributed from the DoD- and NIH-supported Federal Interagency Traumatic Brain Injury Research (FITBIR) Informatics Systems. FITBIR is a collaborative biomedical informatics system created by the Department of Defense and the National Institutes of Health to provide a national resource to support and accelerate research in TBI. Dataset identifiers: FITBIR-STUDY0000220, FITBIR-STUDY0000226, FITBIR-STUDY0000228, FITBIR-STUDY0000234, FITBIR-STUDY0000239, FITBIR-STUDY0000246, FITBIR-STUDY0000256, FITBIR-STUDY0000263, FITBIR-STUDY0000267, FITBIR-STUDY0000279, FITBIR-STUDY0000280, FITBIR-STUDY0000310, FITBIR-STUDY0000313, FITBIR-STUDY0000314, FITBIR-STUDY0000321, FITBIR-STUDY0000329, FITBIR-STUDY0000330, FITBIR-STUDY0000348, FITBIR-STUDY0000349, FITBIR-STUDY0000350, FITBIR-STUDY0000358, FITBIR-STUDY0000363, FITBIR-STUDY0000364, FITBIR-STUDY0000365, FITBIR-STUDY0000409, and FITBIR-STUDY0000411. This manuscript reflects the views of the authors and may not reflect the opinions or views of the DOD, NIH, or of the Submitters submitting original data to FITBIR Informatics System. We used generative AI to debug, edit, and autocomplete code. We take accountability for the review of all work generated by AI.